\documentclass[preprint,12pt, nofootinbib]{revtex4}

\usepackage[brazil, english]{babel}
\usepackage[utf8]{inputenc}
\usepackage{graphicx}
\usepackage{epsfig}
\usepackage{amssymb}
\usepackage{amsmath} 
\usepackage{slashed}
\usepackage[unicode=true,bookmarks=false,breaklinks=false,pdfborder={0 0 1},colorlinks=true]
 {hyperref}
\hypersetup{
 citecolor=blue,linkcolor=blue,urlcolor=blue}

\graphicspath{%
    {converted_graphics/}
    {/}
}

\usepackage{tikz,tikz-3dplot}
\usepackage{orcidlink}
\tdplotsetmaincoords{80}{45}

\tikzset{surface/.style={draw=black, fill=white, fill opacity=.6}}

\usepackage{tikz}
\usetikzlibrary{decorations.pathmorphing,patterns}

\begin{document}

\title{Propagation of tensor perturbation in Horndeski-like gravity}

\author{Fabiano F. Santos\orcidlink{0000-0001-5473-8797}}
\email{fabiano.ffs23-at-gmail.com} \affiliation{School of Physics, Damghan University, Damghan, 3671641167, Iran.}

\author{Jackson Levi Said\orcidlink{0000-0002-7835-4365}}
\email{jackson.said@um.edu.mt} \affiliation{Institute of Space Sciences and Astronomy, University of Malta, Msida, MSD 2080, Malta and Department of Physics, University of Malta, Msida, MSD 2080, Malta.}

\begin{abstract}
Scalar-tensor theories have shown promise in many sectors of cosmology. However, recent constraints from the speed of gravitational waves have put severe limits on the breadth of models such classes of theories can realize. In this work, we explore the possibility of a Horndeski Lagrangian that is equipped with two dilaton fields. The evolution of a two-dilation coupled cosmology is not well-known in the literature. We explore the tensor perturbations in order to assess the behavior of the model again the speed of gravitational wave constraint. Our main result is that this model exhibits of a class of cosmological theories that is consistent with this observational constraint.

 \vskip 3mm
 \normalsize\noindent{\narrower{keywords: Horndeski gravity; Cosmology sector; Tensor perturbations; Observational constraint; Dilaton fields.}}
\end{abstract}

\maketitle
\newpage

\section{Introduction}

Over a century ago, Einstein's General Theory of Relativity provided a revolution in our understanding of gravity and spacetime, laying the foundation for modern cosmology \cite{Haco:2018ske,Haco:2019ggi,Haco:2019paj}. Since its inception, the theory has led to numerous groundbreaking discoveries, including the development of the standard cosmological model, which emerged in the latter half of the 20th century. This model is rooted in two pivotal observations: the recession of galaxies, first identified by Edwin Hubble \cite{Hubble}, and the discovery of the Cosmic Microwave Background (CMB) radiation by Penzias and Wilson \cite{Penzias:1965wn}. These observations provided compelling evidence for the Big Bang paradigm and established a framework for understanding the large-scale structure and evolution of the universe.


Recent advances in observational cosmology, particularly in the study of the CMB and gravitational waves (GWs), have opened new avenues for testing these theoretical frameworks. For instance, the stochastic background of primordial gravitational radiation predicted by cosmology has been shown to be consistent with Planck data \cite{Planck:2018vyg,Planck:2018jri} and may fall within the sensitivity range of current and future GW detectors such as LIGO, Virgo, and LISA \cite{Auclair:2019wcv}. Notably, pre-Big Bang models, as described in \cite{Gasperini:2017qvr}, occupy a wide region of parameter space that allows for the generation of a tensor perturbation spectrum compatible with these observations.


In addition to Einstein's theory, alternative gravitational frameworks such as Lovelock gravity, Horndeski gravity, and Gauss-Bonnet gravity have found applications in cosmology \cite{Sberna:2017nzp, Papallo:2018wuo, Santos:2022lxj, Santos:2023eqp, Santos:2021guj, Santos:2020xox, Santos:2019ljs, Harko:2016xip,Horndeski:2024sjk}. These models are particularly attractive for inflationary scenarios, as they naturally incorporate scalar fields arising from the gravitational sector. Recent precision measurements from the Dark Energy Survey (DES) and Planck satellite observations \cite{Planck:2018vyg} have refined our understanding of this expansion, revealing subtle tensions in the Hubble constant ($H_{0}$) when derived from early versus late universe probes. These discrepancies potentially signal physics beyond the standard $\Lambda$CDM cosmological model, possibly involving modified dark energy equations of state \cite{Riess:2021jrx,DES:2021wwk} or additional relativistic species in the early universe. In this work, we propose a novel cosmological background within the framework of Horndeski-like gravity, focusing on the direct inflationary production of gravitational radiation. Our approach begins with the amplification of tensor metric fluctuations, leading to the formation of a cosmic background of relic gravitons. This setup provides a promising avenue for exploring the interplay between modified gravity theories and the observational signatures of early-universe physics.

Alternative gravitational frameworks, such as Lovelock gravity, Horndeski gravity, and Gauss-Bonnet gravity \cite{Sberna:2017nzp, Papallo:2018wuo, Santos:2022lxj, Santos:2023eqp, Santos:2021guj, Santos:2020xox, Santos:2019ljs, Harko:2016xip,Horndeski:2024sjk}, have emerged as compelling extensions to Einstein's general relativity, offering new insights in cosmology. These theories are particularly appealing in the context of early-universe physics, as they naturally accommodate scalar fields that arise from the gravitational sector, providing a robust theoretical foundation for inflationary models. Unlike standard inflationary scenarios, where scalar fields are often introduced ad hoc, these frameworks integrate them as intrinsic components of the modified gravitational dynamics.

In this work, we explore a novel cosmological background within the framework of Horndeski-like gravity, focusing on its implications for the direct inflationary production of gravitational waves. Specifically, we investigate the amplification of tensor metric fluctuations during inflation, which leads to the generation of a stochastic background of relic gravitons. This mechanism not only enriches our understanding of the interplay between modified gravity and early-universe physics but also offers a testable prediction for future gravitational wave observatories, such as LISA and LIGO \cite{Auclair:2019wcv}. By leveraging the unique features of Horndeski-like gravity, our approach provides a promising avenue for probing the observational signatures of alternative gravitational theories and their role in shaping the dynamics of the primordial universe.


Recently, applications of the Horndeski gravity in cosmology \cite{Santos:2022lxj,Santos:2023eqp,Santos:2021guj,Santos:2020xox,Santos:2019ljs} has been achieved through the following Lagrangian 
\begin{eqnarray}
    {\cal L}_{\rm H}=(R-2\Lambda) -\frac{1}{2}(\alpha g_{\mu\nu}-\gamma\,  G_{\mu\nu})\nabla^{\mu}\phi\nabla^{\nu}\phi\,,\label{L1}
\end{eqnarray}
where $R$, $G_{\mu \nu}$ and $\Lambda$ are the scalar curvature, the Einstein tensor, and the cosmological constant respectively, and where $\phi=\phi(r)$ is a scalar field. Also, $\alpha$ and $\gamma$ are coupling constants. Using the Lagrangian in Ref.~\eqref{L1} equipped with two dilatons, we proposed the scenario of the propagation of tensor perturbations considering the action 
\begin{eqnarray}
    S = -\frac{1}{16\pi\,G}\int{d^{4}x\sqrt{|g|}e^{-\phi}({\cal L}_{\rm H}-\frac{1}{2}\nabla_{\mu}\chi\nabla^{\mu}\chi+V(\phi,\chi))}\,.\label{L2}
\end{eqnarray}
To reproduce the standard Einstein-Horndeski-like equations, we need to add the boundary term $S_{\Sigma}$:
\begin{eqnarray}
    &&S_{\Sigma}=-\frac{1}{8\pi\,G}\int{d^{3}x\sqrt{|g|}e^{-\phi }({\cal L}_{\rm H,\Sigma}+\mathcal{L}_{ct})},\label{L3}\\
    &&\mathcal{L}_{\rm H,\Sigma}=(K-\Sigma)-\frac{\gamma}{4}\nabla_{\mu}\phi\nabla_{\nu}\phi K^{\mu\nu}\nonumber\\
    &&-\frac{\gamma}{4}(\nabla_{\mu}\phi\nabla_{\nu}\phi n^{\mu}n^{\nu}-(\nabla \phi)^2)K,\\
    &&\mathcal{L}_{ct}=c_{0}+c_{1}R+c_{2}R^{ij}R_{ij}\nonumber\\
    &&+c_{3}R^{2}+b_{1}(\partial_{i}\phi\partial^{i}\phi)^{2}+\cdots.
\end{eqnarray}
Here $K_{\mu\nu}=g^{\phantom{\mu}\beta}_{\mu}\nabla_{\beta}n_{\nu}$ is the extrinsic curvature, $K=g^{\mu\nu}K_{\mu\nu}$ is the trace of the extrinsic curvature, $n^{\mu}$ is an outward pointing unit normal vector to the boundary, $\Sigma$ is the boundary tension. ${\cal L}_{ct}$ is the boundary counterterms.

In this work, we will address the decelerated expansion (i.e., from inflationary to standard evolution), which causes an amplification of the quantum fluctuations of the various background fields (the background fields in our case will be the gravity fields Horndeski) and can produce a large number of various types of radiation \cite{Gasperini:2017qvr,Sberna:2017nzp}. The spectral properties of this radiation are strongly correlated with the kinematics of the inflationary phase (or more precisely in our work, with the Horndeski kinetic term of gravity) \cite{Gasperini:2017qvr}. Direct (or indirect) observations of such a primordial component of cosmic radiation could thus provide us with important information about inflationary dynamics, testing the predictions of the various cosmological scenarios.

\section{The Overarching Theory}\label{sec:EoM}

In this section, we develop a framework for inflationary kinematics within the context of Horndeski gravity, a generalized scalar-tensor theory that extends the standard cosmological paradigm (see, e.g., \cite{Horndeski:2024sjk}). Inflationary dynamics in this framework can exhibit deviations from conventional field-theoretic models, particularly in scenarios inspired by cosmology. These deviations are critical for identifying unique signatures inspired by models and understanding their implications for early-universe physics, including the generation of primordial perturbations and their imprints on the large-scale structure (see, e.g., \cite{Byrnes:2006fr, Linde:1990flp}). The effective action for the gravitational sector in our Horndeski-like framework is given by:

\begin{eqnarray}
S_{\text{total}} = S + S_{\Sigma} + S_{m},\label{S1}
\end{eqnarray}
where \( S_{m} \) represents the matter action. By varying this action (\ref{S1}), we derive the Einstein-Horndeski equations of motion:

\begin{eqnarray}
G_{\mu\nu} + \Lambda g_{\mu\nu} = 2T_{\mu\nu},\label{EH1}
\end{eqnarray}
where the energy-momentum tensor \( T_{\mu\nu} \) includes contributions from scalar fields and matter sources. Specifically, \( T_{\mu\nu} \) is expressed as:

\begin{eqnarray}
T_{\mu\nu} = \alpha T^{(1)}_{\mu\nu} + \gamma T^{(2)}_{\mu\nu} + T^{(3)}_{\mu\nu} - g_{\mu\nu}V(\phi, \chi),
\end{eqnarray}

with \( T^{(1)}_{\mu\nu} \), \( T^{(2)}_{\mu\nu} \), and \( T^{(3)}_{\mu\nu} \) given by:
\begin{eqnarray}
    &&T^{(1)}_{\mu\nu}=\nabla_{\mu}\phi\nabla_{\nu}\phi-\frac{1}{2}g_{\mu\nu}\nabla_{\lambda}\phi\nabla^{\lambda}\phi+\nonumber\\
    &&\nabla_{\mu}\chi\nabla_{\nu}\chi-\frac{1}{2}g_{\mu\nu}\nabla_{\lambda}\chi\nabla^{\lambda}\chi,\nonumber\\
    &&T^{(2)}_{\mu\nu}=\frac{1}{2}\nabla_{\mu}\phi\nabla_{\nu}\phi R-2\nabla_{\lambda}\phi\nabla_{(\mu}\phi R^{\lambda}_{\nu)}-\nabla^{\lambda}\phi\nabla^{\rho}\phi R_{\mu\lambda\nu\rho}\nonumber\\
    &&-g_{\mu\nu}\left[-\frac{1}{2}(\nabla^{\lambda}\nabla^{\rho}\phi)(\nabla_{\lambda}\nabla_{\rho}\phi)+\frac{1}{2}(\Box\phi)^{2}-(\nabla_{\lambda}\phi\nabla_{\rho}\phi)R^{\lambda\rho}\right]\nonumber\\
    &&-(\nabla_{\mu}\nabla^{\lambda}\phi)(\nabla_{\nu}\nabla_{\lambda}\phi)+(\nabla_{\mu}\nabla_{\nu}\phi)\Box\phi+\frac{1}{2}G_{\mu\nu}(\nabla\phi)^{2},\nonumber\\
    &&T^{(3)}_{\mu\nu} =e^{\phi}T^{(m)}_{\mu\nu}\,.\nonumber
\end{eqnarray}
Describing the kinematic and geometric contributions of the scalar fields \( \phi \) and \( \chi \). These terms encapsulate the non-trivial interactions between the scalar fields and the spacetime curvature, as well as the coupling to matter sources. For instance, the scalar charge density \( \sigma \) of the sources is defined via the variation of the matter action with respect to \( \phi \):

\begin{eqnarray}
\delta_{\phi} S_{m} = -\frac{1}{2} \int d^4x \sqrt{|g|} \, \sigma \delta \phi.
\end{eqnarray}

\begin{figure}[!ht]
\begin{center}
\includegraphics[width=\textwidth]{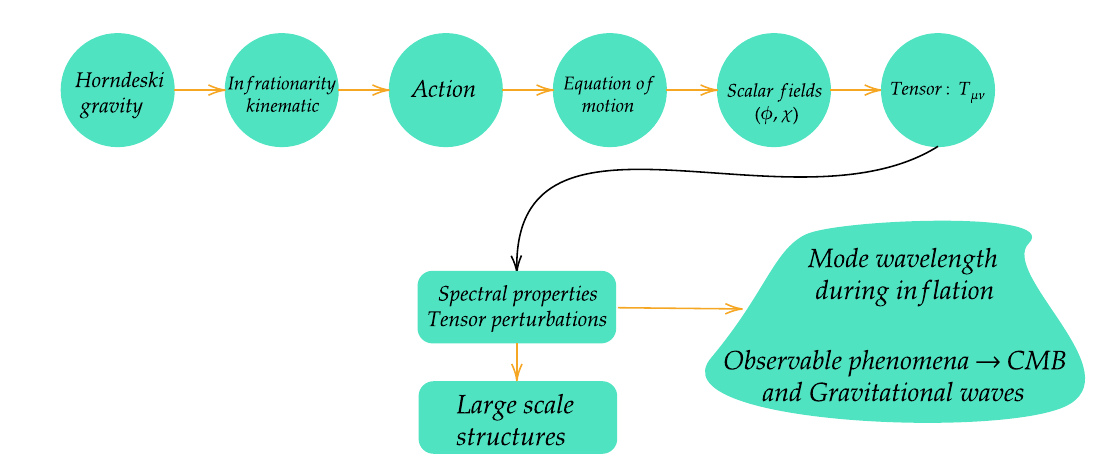}
\end{center}
\caption{The diagram has been successfully constructed and visualized. It illustrates the connections between Horndeski gravity, inflationary kinematics, and their observational implications, including primordial perturbations.}\label{HW}
\label{planohwkhz}
\end{figure}

This formalism allows us to explore potential modifications to the standard cosmological model, particularly in the context of inflationary dynamics and their observational consequences. By connecting the theoretical predictions of Horndeski gravity to observable phenomena, such as the spectral index of primordial perturbations Fig \ref{HW}, we aim to provide a comprehensive framework for testing these models against current and future cosmological data (e.g., \cite{Planck:2018jri}).

In order to study our configuration presented in the diagram Fig \ref{HW} at greater depth, we start with the flat Friedmann–Lema\^{i}tre–Robertson–Walker (FLRW) metric of the form
\begin{equation}
    ds^{2}=-dt^{2}+a^{2}(t)\delta_{ij}dx^{i}dx^{j}\,,\label{13}
\end{equation}
where $a(t)$ is the scale factor, and where the scalar field depends on cosmic time only. With this in hand, we find that the Friedmann equations using the Einstein-Horndeski equations of motion (\ref{EH1}) in the form
\begin{equation}
    H^{2}=\frac{4\kappa\Lambda+\alpha\dot{\phi}^{2}(t)+\dot{\chi}^{2}(t)-2\sigma\exp(\phi)+2V(\phi,\chi)}{3(4\kappa-3\gamma\dot{\phi}^{2}(t))}\,,\label{fri1}
\end{equation}
and
\begin{eqnarray}
    &&(-4\kappa+\gamma\dot{\phi}^{2}(t))H^{2}+4\gamma\,H\dot{\phi}(t)\ddot{\phi}(t)+2\frac{\ddot{a}(t)}{a(t)}(-4\kappa+\gamma\dot{\phi}^{2}(t))+\nonumber\\
    &&(4\kappa\Lambda-2\sigma\exp(\phi)+2V(\phi,\chi)-\dot{\chi}^{2}(t)-\alpha\dot{\phi}^{2}(t)) = 0\,.\label{fri2}
\end{eqnarray}
The equations of motion of the scalar field given an FLRW background are written as
\begin{eqnarray}
    \ddot{\phi}+3H\dot{\phi}+\frac{6\gamma\dot{\phi} H\dot{H}}{\alpha+3\gamma H^{2}}+\frac{1}{\alpha+3\gamma H^{2}}\frac{dV(\phi,\chi)}{d\phi}=0,\label{fri3}\\
    \ddot{\chi}(u)+3H\dot{\chi}+\frac{dV(\phi,\chi)}{d\chi}=0\,,
\end{eqnarray}
where we recovery the usual results of Ref.~\cite{Santos:2019ljs,Rinaldi:2016oqp} for $V(\phi,\chi)=constant$ and $\chi=0$. The motivation to consider these two scalar fields, in particular the field $\chi$ is to find analytical solutions using the first-order formalism approach. With this in hand, we can reduce the equations of motion that are of second order, that is
\begin{eqnarray}
    \dot{\phi}=-W_{\phi}, \nonumber\\
    \dot{\chi}=-W_{\chi},\label{12-first}\\
    H = W \,.\nonumber
\end{eqnarray}
Employing $H=W$, we can explicitly demonstrate that Hubble’s parameter evolves over time as a function of \( W[\phi(t)] \). This formulation not only provides a deeper understanding of the time dependence of Hubble’s parameter but also offers a more systematic framework for analyzing its behavior. Thus, firstly, combining Eq.~\eqref{fri1} with Eq.~\eqref{fri2} for simplicity, we find
\begin{equation}
    WW_{\phi\phi}+\frac{W^{2}_{\phi}}{2}+\frac{4}{3}W^{2}-\beta-2\frac{W^{2}_{\chi}}{\gamma W^{2}_{\phi}} = 0\,,\label{13}
\end{equation}
and now combining the equations Eq.~\eqref{fri1} with Eq.~\eqref{fri3}, we have
\begin{equation}
    WW_{\phi\phi}+\frac{W^{2}_{\phi}}{2}+\frac{4}{3}W^{2}-\beta-\frac{3}{4\gamma}\frac{W_{\chi}W_{\chi\phi}}{WW_{\phi}}=0,\label{14}
\end{equation}
where $\beta=(4\kappa-\alpha)/\gamma$. However, we can see that for consistency when we compare the Eqs.~(\ref{13}--\ref{14}), we get the following constraint on the superpotential 
\begin{equation}
    \frac{3}{8}\frac{W_{\chi\phi}}{W}=\frac{W_{\chi}}{W_{\phi}}\,.\label{15}
\end{equation}
Following \cite{Santos:2023eqp}, we can choosing the simplest superpotential $W(\phi,\chi) = e^{\sqrt{\frac{8}{3}}(\phi+\chi)}$ where this solution is satisfy by the Eq.~\eqref{12-first} which is consistent with Eq.~\eqref{15} given by  
\begin{eqnarray}
    &&\phi(t)=\frac{3}{8}\ln(t),\label{16}\\
    &&\chi(t)=\frac{3}{8}\ln(t),\label{17}\\
    &&H(t)=\frac{9}{16}\frac{1}{t}.\label{18}\\
    &&a(t)=a_{0}t^{9/16}\,,\label{19}
\end{eqnarray}
which is the general solution of the system. Such solution can be found thanks to the field $\chi$, which provides an analytical system, the other hand, if $\chi=0$ with $\phi\neq\,0$, according to equation (\ref{15}) $W(\phi,\chi)=$constant. Of course in the regime $\chi\to\,0$, we recover some description of cosmological inflation in Einstein-Horndeski gravity, but performed by numerical computation \cite{Santos:2021guj,Santos:2019ljs,Rinaldi:2016oqp}; the idea is to provide an analytical description of the Horndeski gravity \cite{Harko:2016xip}. Besides, the advantage of our model is that we do not choose the potential, this emerges naturally from first-order formalism with the scalar field $\chi$.

\section{Propagation of tensor perturbations}
We now analyze the tensor perturbations, which encapsulate the spectral properties of radiation in the context of cosmological evolution. Following the methodologies outlined in Refs.~\cite{Santos:2022lxj, Santos:2023eqp, Santos:2021guj, Brito:2018pwe}, we employ the TT (transverse and traceless) gauge, where $\partial^{\mu}h_{\mu\nu} = 0$ and $h^{\mu}_{\mu} = 0$, with the scalar field perturbation $\delta\phi$ set to zero. Under these conditions, the equations governing tensor fluctuations can be derived. These tensor modes are directly linked to observable phenomena in the recent universe, such as the polarization patterns in the Cosmic Microwave Background (CMB) and the gravitational wave background. By studying these perturbations, we gain insights into the early universe's dynamics and the imprints left on the large-scale structure of the cosmos. 

Tensor perturbations are crucial for understanding the universe's evolution. They manifest as gravitational waves, which can be detected through their influence on the CMB polarization \cite{Namikawa:2021obu} or through direct observations by gravitational wave detectors like LIGO, Virgo, and future missions such as LISA \cite{Auclair:2019wcv}. These observations provide a window into the physics of the early universe, including inflationary models and potential deviations from standard cosmology. The equations for the tensor fluctuations given by

\begin{eqnarray}
    &&T(t)\ddot h_{\mu\nu}+B(t)\dot{h}_{\mu\nu}-a^{-2}\nabla^2h_{\mu\nu}=0,\label{Ten1}\\
    &&T(t)=\frac{4\kappa-\gamma\dot{\phi}^2}{4\kappa+\gamma\dot{\phi}^2},\\
    &&B(t)=\frac{3H(4\kappa-\gamma\dot{\phi}^2)-2\gamma\dot{\phi}\ddot{\phi}-4\kappa\dot{\phi}}{4\kappa+\gamma\dot{\phi}^2}\,.
\end{eqnarray}
The expression in Eq.~\ref{Ten1} has a similar form as in Refs.~\cite{Santos:2021guj} except for the coefficients of $T(t)$ and $B(t)$. In the linear regime, the components of $h_{\mu\nu}$ can be decoupled and satisfy an evolution equation in the Einstein-Horndeski frame. We can note that the above equations are obtained from the lowest-order effective action and are valid in the low-energy limit. However, in our case, we only have second-order equations with Horndeski corrections. Another way to derive the expression in Eq.~\eqref{Ten1} is through the ADM formalism \cite{Santos:2021guj}. For this, we can start with the following action
\begin{eqnarray}
    S^{(2)}_{\rm tensor}=\frac{1}{16\pi\,G}\int{dtd^3xa^3e^{-\phi}\left[G_T \dot{h}_{ij}^2-\frac{F_T}{a^2}(\partial_k h_{ij})^2\right]}\,,\label{adm}
\end{eqnarray}
where the TT gauge is given by $\partial^i h_{ij}=0, h_i^i=0$. With this, we can find the equations for the tensor fluctuations as
\begin{eqnarray}
    T(t)\ddot h_{ij}+B(t)\dot{h}_{ij}-a^{-2}\nabla^2 h_{ij}=0\,,\label{T1}
\end{eqnarray}
with
\begin{eqnarray}
    T(t)\equiv\frac{G_T}{F_T}\,, \:\:\: B(t)\equiv\frac{\dot{G}_T+3HG_T}{F_T}\,.
\end{eqnarray}
We can find that the explicit form of the coefficients $G_T$ and $F_T$ reads \cite{Kob}
\begin{eqnarray}
    && G_T\equiv 2\left[G_4-2XG_{4X}-X\left(H\dot\phi G_{5X}-G_{5\phi}\right)\right]\nonumber\\
    &&=2\left(G_4+X G_{5\phi}\right)\,,\\
    && F_T\equiv 2\left[G_4-X\left(\ddot\phi G_{5X}+G_{5\phi}\right)\right]\nonumber\\
    &&=2\left(G_4-X G_{5\phi}\right)\,.
\end{eqnarray}
Here, the coefficients are simplified under our choice of $G_i$'s in the action (\ref{L2}). In this case, we have
\begin{eqnarray}
    T(t)=\frac{4\kappa-\eta\dot{\phi}^2}{4\kappa+\eta\dot{\phi}^2}, \:\:\: B(t)=\frac{3H(4\kappa-\eta\dot{\phi}^2)-2\eta\dot{\phi}\ddot{\phi}}{4\kappa+\eta\dot{\phi}^2}\,.
\end{eqnarray}
The motivation to write the action in Eq.~\eqref{adm} is to find the polarization mode $h_{A}(t,x_\mu)$ as a function of conformal time $d\eta=dt/a$. Under these considerations the action in Eq.~\eqref{adm} becomes
\begin{eqnarray}
    S^{(2)}_{\rm tensor}=\int{d\eta\,d^3xz^{2}(\eta)\left[G_Th^{'2}_{ij}-F_T(\partial_k h_{ij})^2\right]}\,,\label{adm1}
\end{eqnarray}
where
\begin{eqnarray}
    z(\eta)=\frac{ae^{-\phi/2}}{\sqrt{16\pi\,G}}\,,
\end{eqnarray}
describes the dynamics of a scalar field variable $h$ with a non-minimal coupling to a time-dependent background field $z(\eta)$, which also called the ``pump field''. In our cosmic background in Horndeski-like gravity, this field represents a dilaton and an external and internal scale factor (known as "moduli" field). Now, considering the following transformation $u=zh$, we can rewrite the action in Eq.~\eqref{adm1} as
\begin{eqnarray}
    S^{(2)}_{\rm tensor}=\int{d\eta\,d^3x\left[G_Tu^{'2}+F_Tu\nabla^{2}u+F_T\frac{z^{''}}{z}u^{2}\right]}\,,\label{adm}
\end{eqnarray}
the variation of the action (\ref{adm1}) leads to the canonical Schrodinger-like evolution equation as in the form
\begin{eqnarray}
    u^{''}-\frac{F_T}{G_T}(\nabla^{2}+U(\eta))u=0;\quad\,U(\eta)=\frac{z{''}}{z}\,,\label{adm2}
\end{eqnarray}
where the gravitational wave speed is defined in terms of 
\begin{eqnarray}
    c^{2}_{GW}\equiv\frac{F_T}{G_T}\,.\label{adm3}
\end{eqnarray}
Beyond, the effective action in Eq.~\eqref{adm1}
\begin{eqnarray}
    &&S^{(2)}_{\rm tensor}=\int{d\eta\,d^3x\mathcal{L}}\label{adm4}\\
    &&\mathcal{L}=\frac{1}{2}\left[G_Tu^{'2}-F_T(\nabla_iu)^2+F_T\frac{z^{''}}{z}u^{2}\right]\,.
\end{eqnarray}
The field can be quantized starting from the definition of the usual momentum density given by 
\begin{eqnarray}
    \mathcal{P}=\frac{\partial\mathcal{L}}{\partial\,u^{'}}=G_Tu^{'}\,.
\end{eqnarray}

\begin{figure}[!ht]
\begin{center}
\includegraphics[scale=0.14]{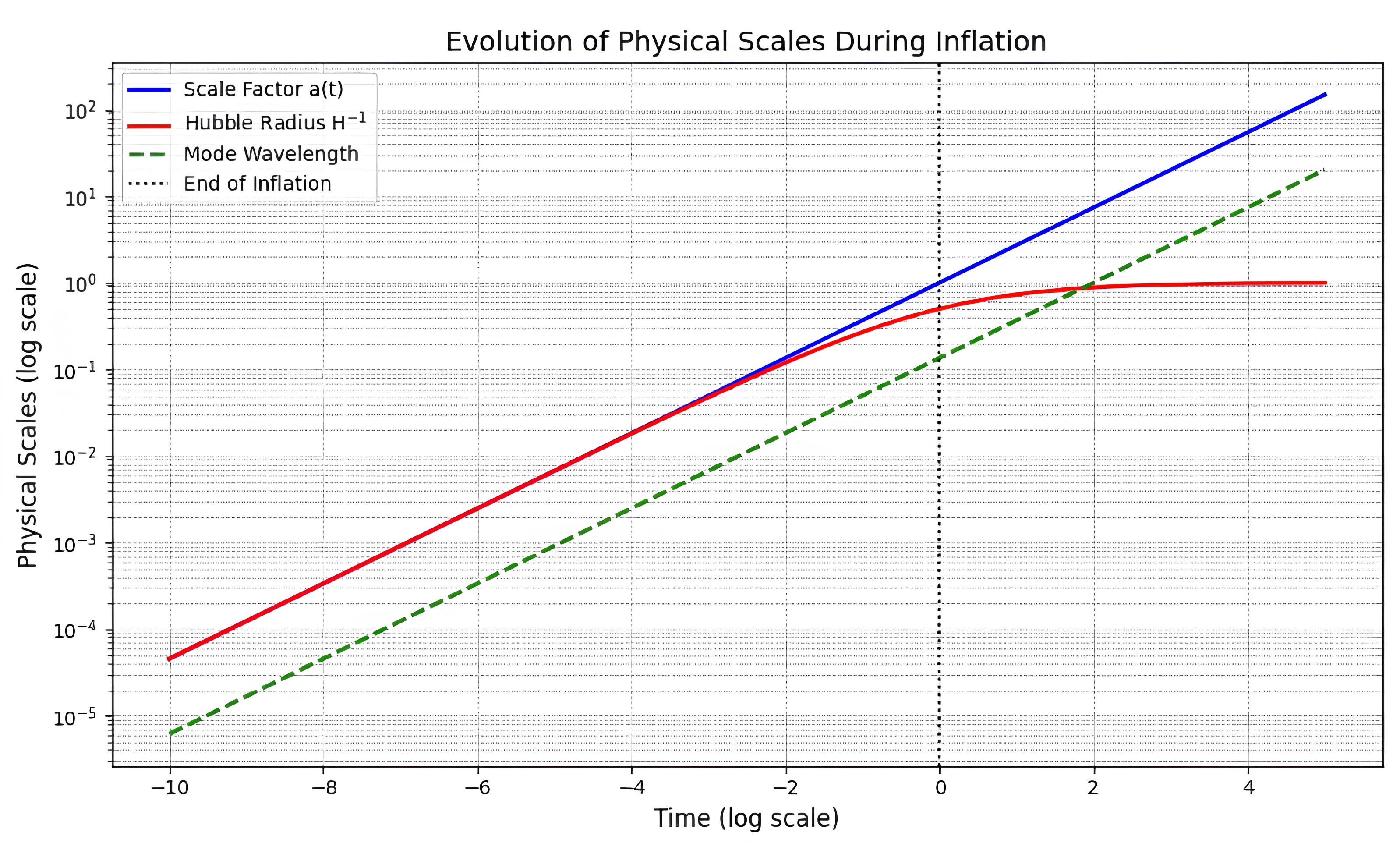}
\end{center}
\caption{The diagram has been constructed and visualizes the relationship between the scale factor, Hubble radius, and mode wavelength during inflation. It shows how the mode's wavelength grows faster than the Hubble radius, eventually leaving the horizon, and re-entering later.}\label{HW1}
\label{planohwkhz}
\end{figure}

Now, we can impose (equal-time) canonical commutation relations
\begin{eqnarray}
    [u(x_i,\eta),\mathcal{P}(x^{'}_{i})]=i\delta^{3}(x-x^{'})\,,
\end{eqnarray}
on the $\eta=constant$ hypersurface. In our prescription, the classical variable $u$ can be promoted to a field operator and, thus, it can be expanded over a complete set of solutions of the classical solutions in Eq.~\ref{adm2}. Through the Fourier modes
\begin{eqnarray}
    u(x_i,\eta)=\int{\frac{d^3k}{(2\pi)^3}[a_ku_ke^{ik_ix^i}+a^{\dagger}_ku^*_ke^{-ik_ix^i}]}\,,
\end{eqnarray}
we can write the eigenvalue equation
\begin{eqnarray}
    u^{''}_{k}+c^{2}_{GW}(k^{2}+U(\eta))u_{k}=0;\quad\,U(\eta)=\frac{z{''}}{z}\,.\label{adm5}
\end{eqnarray}
Using the Fourier modes, we can provide the commutation relation as
\begin{eqnarray}
    &&G_T[a_{k},a_{k'}]=G_T[a^{\dagger}_{k},a^{\dagger}_{k'}]=0\\
    &&G_T[a_{k},a^{\dagger}_{k'}]=\delta^{3}(k-k')\,,\label{adm6}
\end{eqnarray}
where we can apply the usual interpretation with $a_k$ annihilation operator and $a^{\dagger}_{k}$ creation operator. The normalization condition is given by
\begin{eqnarray}
    u_ku^{'*}_k-u^{'}_ku^{*}_k=iG_T\,,\label{adm7}
\end{eqnarray}
with $G_T=4\kappa-\gamma\,a^{-2}\phi^{'2}=4\kappa-\gamma/(a^2_0\eta^2)$. The orthonormality is given by the scalar product
\begin{eqnarray}
    \langle\bar{u}_k|\bar{u}_{k'}\rangle\equiv-i\int{d^3x(\bar{u}_k\bar{u}^{'*}_{k'}-\bar{u}^{'}_k\bar{u}^{*}_{k'})}=\delta^3(k-k')\,.\label{adm8}
\end{eqnarray}
The normalization condition in Eq.~\ref{adm7} can now be applied to fix the initial amplitude of the metric's quantum fluctuations in a typical inflationary scenario. For the asymptotic regime with $\eta\to-\infty$, we have that the $G_T=4\kappa$, i.e., we recover the usual case with $\kappa=1/4$. Besides, using the equations of $\phi(t)$ and $a(t)$ rewrite in terms of $\eta$ by $d\eta=dt/a$, we can give the form of $z$ as $z=a_0/\sqrt{16\pi\,G}$, provide $U(\eta)=0$ and we have a free field equation  
\begin{eqnarray}
    u^{''}_{k}+k^{2}c^{2}_{GW}u_{k}=0\,.\label{adm9}
\end{eqnarray}
In this case, the expression in Eq.~\eqref{adm9} has oscillating solutions. In the inflationary regime, the solution exhibits a positive frequency mode on the initial hypersurface. This behavior arises because the scale factor of the Universe, \(a(t)\), grows exponentially during inflation, outpacing the growth of the Hubble radius, \(H^{-1}\) Fig \ref{HW1}. Consequently, the physical wavelength of each mode stretches faster than the Hubble radius, leading to a scenario where the mode's wavelength eventually exceeds the Hubble radius—a process commonly referred to as the mode "leaving the horizon." This phenomenon is critical for understanding the generation of primordial perturbations, as modes that exit the horizon during inflation become "frozen" and later re-enter during the radiation or matter-dominated eras, leaving observable imprints on the Cosmic Microwave Background (CMB) and the large-scale structure of the Universe. Once inflation ends, the dynamics of these modes are governed by the post-inflationary evolution of the Universe \cite{Armendariz-Picon:2008pwa,Flanagan:2005yc,Kubota:2022lbn}.

\begin{figure}[!ht]
\begin{center}
\includegraphics[scale=0.14]{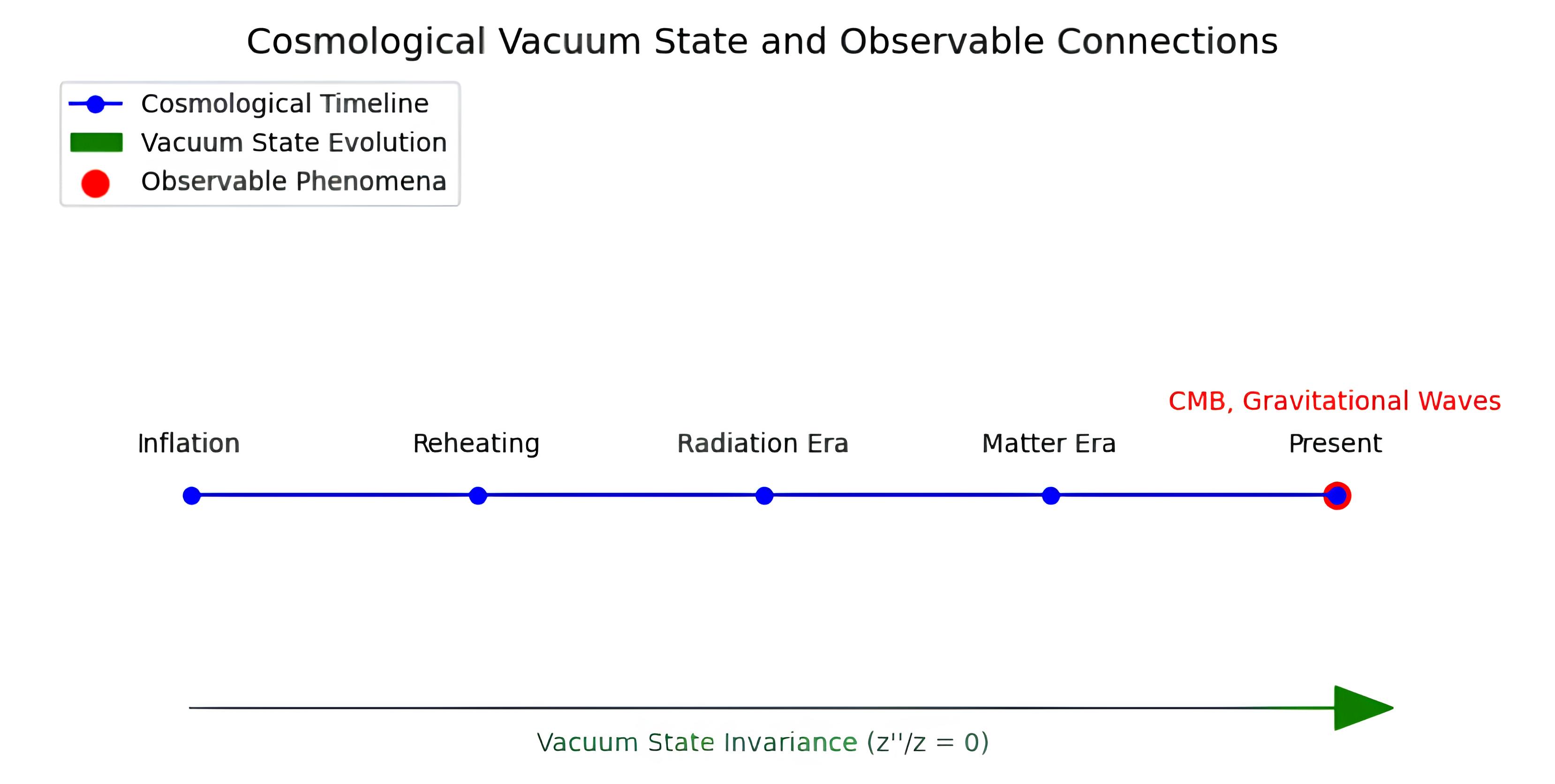}
\end{center}
\caption{The diagram illustrating the cosmological timeline, the invariance of the vacuum state, and its connection to observable phenomena like the CMB and gravitational waves.}\label{HW2}
\label{planohwkhz}
\end{figure}

The general solution for equation (\ref{adm9}) is given by
\begin{eqnarray}
    u_k=\frac{e^{ik^{2}c^{2}_{GW}\eta}}{\sqrt{kc_{GW}}}\,,\label{adm10}
\end{eqnarray}
the gravitational wave speed in the regime $\eta\to-\infty$, becomes 
\begin{eqnarray}\label{eq:speed_of_prop}
    c^{2}_{GW}\equiv\frac{F_T}{G_T}=\frac{4\kappa+\gamma/(a^2_0\eta^2)}{4\kappa-\gamma/(a^2_0\eta^2)}\to1=c^{2}_{light}\,.
\end{eqnarray}
In the context of early universe cosmology, the normalized positive frequency solution in Eq.~\eqref{adm10} plays a pivotal role in defining the initial vacuum state of tensor fluctuations. This vacuum state corresponds to the lowest-energy configuration associated with the action in Eq.~\eqref{adm1}. For a time-dependent cosmological background, the effective mass term, given by \( z''/z = 0 \), ensures that the notion of positive frequency remains invariant over time. As a result, the vacuum state is uniquely defined and globally consistent across the entire cosmological manifold. This property implies that a state initially devoid of particles (i.e., quanta of tensor fluctuations) remains empty throughout the evolution of the universe. Such a framework is critical for understanding the quantum origins of cosmological perturbations and their imprint on observable phenomena, such as the Cosmic Microwave Background (CMB) anisotropies and primordial gravitational waves (see, e.g., \cite{Hu:1995kot,Page:1993mn}). These insights provide a foundation for connecting theoretical predictions with empirical observations Fig \ref{HW2}, thereby advancing our understanding of the universe's earliest moments.

\section{Conclusions and discussions}

The constraints set on the speed of propagation of gravitational waves has turned out to be a stringent constraint on general classes of scalar-tensor theories. In the present work, we explore the possibility of a two-dilaton field scenario. Such scenarios have become interesting in recent years with the appearance of problems such as the cosmic tensions issue. Here, the scalar fields may offer a dual action in which one has an action in the early Universe while the other is free to act more globally over time. In our case, we do not constrain these effects but we do explore the tensor perturbations of this setting with a focus on assessing where the class of theories is constrained in any way by the speed constraint on gravitational wave propagation.

Our Horndeski-like model is defined in Eq.~\eqref{S1}, which contains a two-dilaton action~\eqref{L2} together with the regular matter action $S_{m}$ and a boundary component~\eqref{L3} that is important for preserving a healthy general relativistic limit~\eqref{L3}. The eventual equations of motion are then derived in Sec.~\ref{sec:EoM} where the dynamical variables are expressed at background level through the FLRW metric. Indeed, as an example, we provide a particular solution for one simple expression of the superpotential that describes the two-dilaton fields. Ultimately, for this case, we find the simple solution expressed in Eqs.(\ref{16}--\ref{19}).

The propagation of tensor perturbations in this framework is governed by modified equations of motion that incorporate the effects of the two dilaton fields. These modifications are consistent with the stringent constraints on the speed of gravitational waves, as established by recent multi-messenger observations. The model's ability to satisfy these constraints while simultaneously addressing the $H_0$ tension highlights its robustness and viability as an alternative to the standard $\Lambda$CDM paradigm. The unique features of the Horndeski-like gravity framework, including its intrinsic scalar fields and modified gravitational dynamics, offer a promising avenue for probing the observational signatures of alternative cosmological models. In conclusion, the two-dilaton model within the Horndeski-like gravity framework represents a significant step forward in addressing the $H_0$ discrepancy and advancing our understanding of the early universe. By leveraging the interplay between tensor perturbations and modified gravity, this approach provides a comprehensive framework for testing the predictions of alternative cosmological theories against current and future observational data. Future work should focus on refining the model's parameters and exploring its implications for other cosmological tensions, such as those related to dark energy and the large-scale structure of the universe.

The cosmic expansion history, characterized by the Hubble parameter evolution discussed in our diagram Fig \ref{HW2}, fundamentally connects with gravitational wave propagation through the underlying spacetime geometry. Tensor perturbations, representing the spin-2 field disturbances of the FLRW metric, provide a complementary probe to the scalar sector measurements that drive the Hubble tension \cite{Perivolaropoulos:2021jda}. Our analysis in the transverse-traceless (TT) gauge yields the tensor perturbation propagation equation shown in Eq.\eqref{T1}, which reveals how gravitational waves traverse the expanding cosmos. Remarkably, the propagation speed derived in Eq.\eqref{adm3} approaches the speed of light in its asymptotic limit as expressed in Eq.~\eqref{eq:speed_of_prop}, aligning with the stringent constraints from GW170817 and its electromagnetic counterpart \cite{LIGOScientific:2017zic}. This consistency strengthens our modified gravity framework while simultaneously addressing the $H_{0}$ discrepancy illustrated in our analysis. The tension between early and late universe measurements may find resolution in the tensor sector's behavior, complementing potential explanations from early dark energy models \cite{Smith:2019ihp} or running vacuum scenarios \cite{Givans:2020sez}. Our future investigations will extend to scalar perturbations, which govern structure formation and could impose additional constraints on viable cosmological models, potentially narrowing the parameter space where Hubble tension solutions may exist.

The study of tensor perturbations in modified gravity frameworks, such as Horndeski-like gravity, has become increasingly significant in addressing key cosmological challenges. Tensor modes, as spin-2 perturbations of the metric, provide a unique observational window into the dynamics of the early universe, particularly through their imprints on the Cosmic Microwave Background (CMB) and gravitational wave observatories. These modes are instrumental in constraining viable cosmological models, as they directly probe the interplay between the gravitational sector and the underlying spacetime geometry.

In future work, we will address the integration of viscous fluid cosmology into the Horndeski-like framework represents a promising direction for future research. By bridging the gap between modified gravity theories and effective fluid dynamics, this approach has the potential to address some of the most pressing challenges in modern cosmology, including the nature of dark energy, the origin of cosmic acceleration, and the resolution of the Hubble tension.

\section{Data Availability Statement: No Data associated in the manuscript}

$\bullet$The datasets generated during and/or analysed during the current study are available in the arxiv repository,
https://arxiv.org/abs/2401.03558.

$\bullet$ The datasets generated during and/or analysed during the current study are available from the corresponding
author on reasonable request.

$\bullet$ All data generated or analysed during this study are included in this published article (and its supplementary
information files).

$\bullet$ The datasets generated during and/or analysed during the current study are publicly available.
Data sharing not applicable to this article as no datasets were generated or analysed during the current study

\section*{Declarations}

$\bullet$ This manuscript is available in the repository https://arxiv.org/abs/2401.03558.

$\bullet$ Have no competing interests between the authors.

$\bullet$ The author contributed equally to this manuscript.

\section*{Data availability statement}

$\bullet$ The data are in he repository https://arxiv.org/abs/2401.03558.

$\bullet$ All original data for this work can be found at https://arxiv.org/abs/2401.03558.

\section*{Author contributions}

All authors contributed to the study conception and design. Material preparation, data collection and analysis were performed by Fabiano F. Santos and Jackson Levi Said. The first draft of the manuscript was written by  Fabiano F. Santos and Jackson Levi Said commented on previous versions.  


\end{document}